\documentclass[12pt,oneside,letterpaper]{article}
\usepackage[hmargin=1.25in]{geometry}
\usepackage{graphicx,mathptmx,url,xspace,array}
\usepackage[plain]{fancyref}
\usepackage[small]{caption2}
\usepackage[onehalfspacing]{setspace}
\usepackage{natbib}

\bibpunct{(}{)}{,}{}{}{}

\newcommand{\s}{\ensuremath{s}}
\providecommand{\atanh}{\mathrm{atanh}}
\newcommand{\Jres}{\ensuremath{J_{\mathrm{res}}}}
\newcommand{\Jimm}{\ensuremath{J_{\mathrm{imm}}}}
\newcommand{\Jint}{\ensuremath{J_{\mathrm{int}}}}
\newcommand{\mres}{\ensuremath{m_{\mathrm{res}}}}
\newcommand{\mimm}{\ensuremath{m_{\mathrm{imm}}}}
\newcommand{\hres}{\ensuremath{h_{\mathrm{res}}}}
\newcommand{\himm}{\ensuremath{h_{\mathrm{imm}}}}

\begin{document}
\setlength{\parskip}{\bigskipamount}
\doublespacing
\title{Equilibria of culture contact derived from ingroup and outgroup
  attitudes}
\author{Pierluigi
  Contucci\textsuperscript{1,2} \and Ignacio Gallo\textsuperscript{1}
  \and Stefano Ghirlanda\textsuperscript{2,3}}
\date{}
\maketitle
\begin{enumerate}
\item Department of Mathematics, University of Bologna.
\item Stockholm University Centre for the Study of Cultural Evolution.
\item Department of Psychology, University of Bologna.
\end{enumerate}

% \noindent
% \textbf{One-sentence summary:} We introduce a mathematical model of
% culture contact and show that a small fraction of immigrants can have
% a dramatic impact of residents' culture.

%\noindent
%\textbf{Corresponding author:} Pierluigi Contucci, Dipartimento di
%  Matematica, P.za di Porta S. Donato 5, 40126 Bologna, Italy. Tel:
%  +39 051 2094404, Fax: +39 051 2094490, Email: contucci@dm.unibo.it.

\clearpage

\begin{abstract}
  Modern societies feature an increasing contact between cultures, yet
  we have a poor understanding of what the outcomes might be. Here we
  consider a mathematical model of contact between social groups,
  grounded in social psychology and analyzed using tools from
  statistical physics. We use the model to study how a culture might
  be affected by immigration. We find that in some cases residents'
  culture is relatively unchanged, but in other cases residents may
  adopt the opinions and beliefs of immigrants.  The decisive factors
  are each group's cultural legacy and its attitudes towards in- and
  out-groups. The model can also predict how social policies may
  influence the outcome of culture contact.
\end{abstract}

\section{Introduction}

Contact between cultures is a prominent feature of modern society,
driven by large-scale migration, global media, communication networks,
and other socio-economical forces \citep{lull00}. Understanding how
human cultures interact is crucial to such issues as immigration
management and the survival of national and minority cultures
\citep{corn94,give05}, yet the dynamics of culture contact are poorly
known. Here we explore the problem considering a single cultural trait
that can take two forms, such as being in favor or against the death
penalty, or whether to wear or not a particular piece of clothing.  We
are interested in: 1) how the two trait forms are distributed among
subgroups in a population, e.g., residents and immigrants, males and
females, social classes, etc.; and 2) how different subgroups
influence each other's traits. We start with assumptions about how
individuals may change their opinions and behaviors as a consequence
of social interactions, and then derive the population-level
consequences of such assumptions using tools from statistical
physics. We study the case of immigration in detail, and conclude that
culture contact may sometime result in residents taking on the
opinions and beliefs of immigrants, depending on each group's cultural
legacy, its attitudes toward in- and out-groups, and social policies.

Our main assumption about how individuals may change their traits is
the so-called ``similarity-attraction'' hypothesis of social
psychology: an individual tends to agree with those who are perceived
as similar to oneself, and to disagree with those who are perceived as
different \citep{byrne71,grant93,byrn97,mich02}.  Additionally,
individuals can be influenced by other forces favoring one form of a
trait over the other. For instance, a media campaign may advertise in
favor or against a given idea or behavior. To determine the
distribution of opinions among subgroups of a population (after they
have been in contact for some time) we apply statistical mechanics, a
branch of theoretical physics that studies the collective properties
of systems composed of many parts that interact according to given
rules.  These techniques were originally developed to derive the laws
of thermodynamics from molecular dynamics \citep{thompson79}, but have
been applied also to biological \citep{amit89,arbi03} and social
systems, including models of social choice
\citep{schelling73,granovetter78,durl99,watts02,macy_etal03}. The
latter consider how collective opinions or choices emerge within a
homogeneous social group. The model we discuss here is, to our
knowledge, the first one to consider populations consisting of
different social groups in interaction.

\section{General framework}

We now outline a general framework for modelling group-level
consequences of interactions among individuals
\citep{contucci_ghirlanda07}. Individual $i$ is described by a binary
variable $\s_i=\pm1$, representing the two possible forms of the
considered trait ($i=1,\ldots,N$). A group is characterized by the
mean value $m=\frac{1}{N}\sum_i s_i$, which can be measured by, say, a
referendum vote or a survey. To apply statistical mechanics we
formalize as follows the interaction between individuals. Let $L_{ik}$
be the similarity that individual $i$ perceives with $k$ and assume
that $L_0$ is the level of similarity above which individuals tend to
agree, and below which they tend to disagree. We can then recast the
similarity-attraction hypothesis in the form of a minimization rule,
assuming that $i$, when interacting with $k$, tends to assume the
trait value that minimizes the quantity
\begin{equation}
  \label{eq:Hik}
  H_{ik}(\s_i,\s_k)=-(L_{ik}-L_0)\s_i\s_k  
\end{equation}
This assumption agrees with the similarity-attraction hypothesis
because, when $L_{ik}>L_0$ the expression is minimized by agreement
($\s_i\s_k=1$), and when $L_{ik}<L_0$ by disagreement ($\s_i\s_k=-1$).
For simplicity we let $L_{ik}-L_0=J_{ik}$ in the following. Then
$J_{ik}>0$ favors agreement and $J_{ik}<0$ favors disagreement. We say
``favor'' because we do not assume strict minimization. Rather, we
assume that a trait value yielding a lower value of (\ref{eq:Hik})
occurs with a higher probability, but not with certainty (see
appendix).  The rationale for this assumption is that similarity to
others is not the sole determinant of an individual's trait
\citep{byrn97,mich02}.

When an individual interacts with many others, we assume that she
tends to minimize the sum, $H_i$, of all functions (\ref{eq:Hik})
relative to each interaction:
\begin{equation}
  \label{eq:Hi}
  H_i(\s_i)= \sum_k H_{ik}(\s_i,\s_k)=-\sum_k J_{ik}\s_i\s_k
\end{equation}
where the sum extends over all individuals with whom $i$ interacts.
In summary, the effect of changing one's trait is gauged according to
whether it makes an individual agree or disagree with others,

The effect of factors such as the media or social norms is, for each
individual, to favor a particular trait value. This can be included
in our minimization scheme by adding a term to (\ref{eq:Hi}):
\begin{equation}
  \label{eq:Hh}
  H_i(\s_i) = -\sum_{k}J_{ik}\s_i\s_k - h_i\s_i  
\end{equation}
The added term means that $\s_i=1$ is favored if $h_i>0$, while
$h_i<0$ favors $\s_i=-1$.  We now define a group-level function $H$
as the sum of individual functions:
\begin{equation}
  \label{eq:H}
  H(\s) = \sum_i H_i(\s_i) = -\sum_{i,k}J_{ik}\s_i\s_k -\sum_ih_i\s_i
\end{equation}
where $\s=\{\s_1,\ldots,\s_N\}$ is the set of all individual traits.
The function $H$ is referred to as the system \textit{Hamiltonian} in
statistical mechanics, where it usually arises from consideration of a
system's physical properties. Here, on the other hand, we have
designed the function $H$ so that lower values of $H$ correspond to
group states that, given our assumptions about individual psychology,
are more likely to occur.  It is this property that allows us to use
statistical mechanics \citep{thompson79}. Note that we do not assume
that individuals explicitly carry out, or are aware of, the
computations in \fref{eq:H}.

\section{Culture contact in immigration}

We consider a large and a small group, referred to, respectively, as
\textit{residents} ($R$) and \textit{immigrants} ($I$).  We are
interested in the effect of one group upon the other, i.e., how
culture contact changes the mean trait values in the two groups. If
residents and immigrants have markedly different culture, the
similarity-attraction hypothesis implies that $J_{ik}$ should be
positive for interactions within a group and negative for interactions
between groups.  A simple assumption (called {\it mean field} in
statistical mechanics) is that $J_{ik}$ depends only on group
membership. This corresponds to the ingroup and outgroup concepts of
social psychology \citep{brown00} and can be formalized as follows:
\begin{equation}
  \label{eq:J}
  J_{ik} = \left\{
    \begin{array}{l@{\qquad}l}
\displaystyle{ \frac{\Jres}{2N}}>0 & i,k\in R\\ \\
\displaystyle{    \frac{\Jint}{2N}}<0 & i\in R, k\in I,\ \mbox{or}\ i\in I, k\in R\\ \\
  \displaystyle{  \frac{\Jimm}{2N}}>0 & i,k\in I
  \end{array}\right.
\end{equation}
where the factor $1/2N$ guarantees that the group function, equation
(\ref{eq:H}), grows proportionally to the number of individuals.
Before the two groups interact, residents and immigrants are each
characterized by a cultural legacy that includes given mean values of
the considered trait, say $\mres^*$ and $\mimm^*$. Our goal is to
predict the values \mres\ and \mimm\ after the interaction has taken
place. To describe the effect of cultural legacies we reason as
follows. Within a group, a mean trait $m^*\ne0$ means that the two
forms of the trait are not equally common. Thus preexisting culture
can be seen as a force that favors one trait value over the other, and
can be modeled by a force term as in (\ref{eq:Hh}).  In other words, a
model in which individuals are biased so that the mean trait is $m^*$
is equivalent to a model with unbiased individuals subject to a force
$h^*$ of appropriate intensity.  The latter can be calculated by
standard methods of statistical mechanics as
\begin{equation}
  \label{eq:h}
  h^* = \atanh(m^*)-Jm^*
\end{equation}
where $\atanh(\cdot)$ is the inverse hyperbolic tangent and $J$ is the
ingroup attitude \citep{thompson79,contucci_ghirlanda07}. Statistical
mechanics also allows to calculate the values of \mres\ and \mimm\
after culture contact as the solution of a system of equations (see
appendix):
\begin{equation}
  \label{eq:m1m2}
  \left\{
  \begin{array}{rcl}
    \mres &=& \tanh\big((1-\alpha)\Jres\mres+\alpha\Jint\mimm +
    \hres^*\big)\\ 
    \\
    \mimm &=& \tanh\big((1-\alpha)\Jint\mres+\alpha\Jimm\mimm +
    \himm^*\big)  
  \end{array}\right.
\end{equation}
where $\tanh(\cdot)$ is the hyperbolic tangent, $\alpha$ is the
fraction of immigrants in the compound group, and $\hres^*$ and
$\himm^*$ are calculated for each group according to (\ref{eq:h}).
\begin{figure}[tp]
  \centering
  \includegraphics[angle=-90,width=\textwidth]{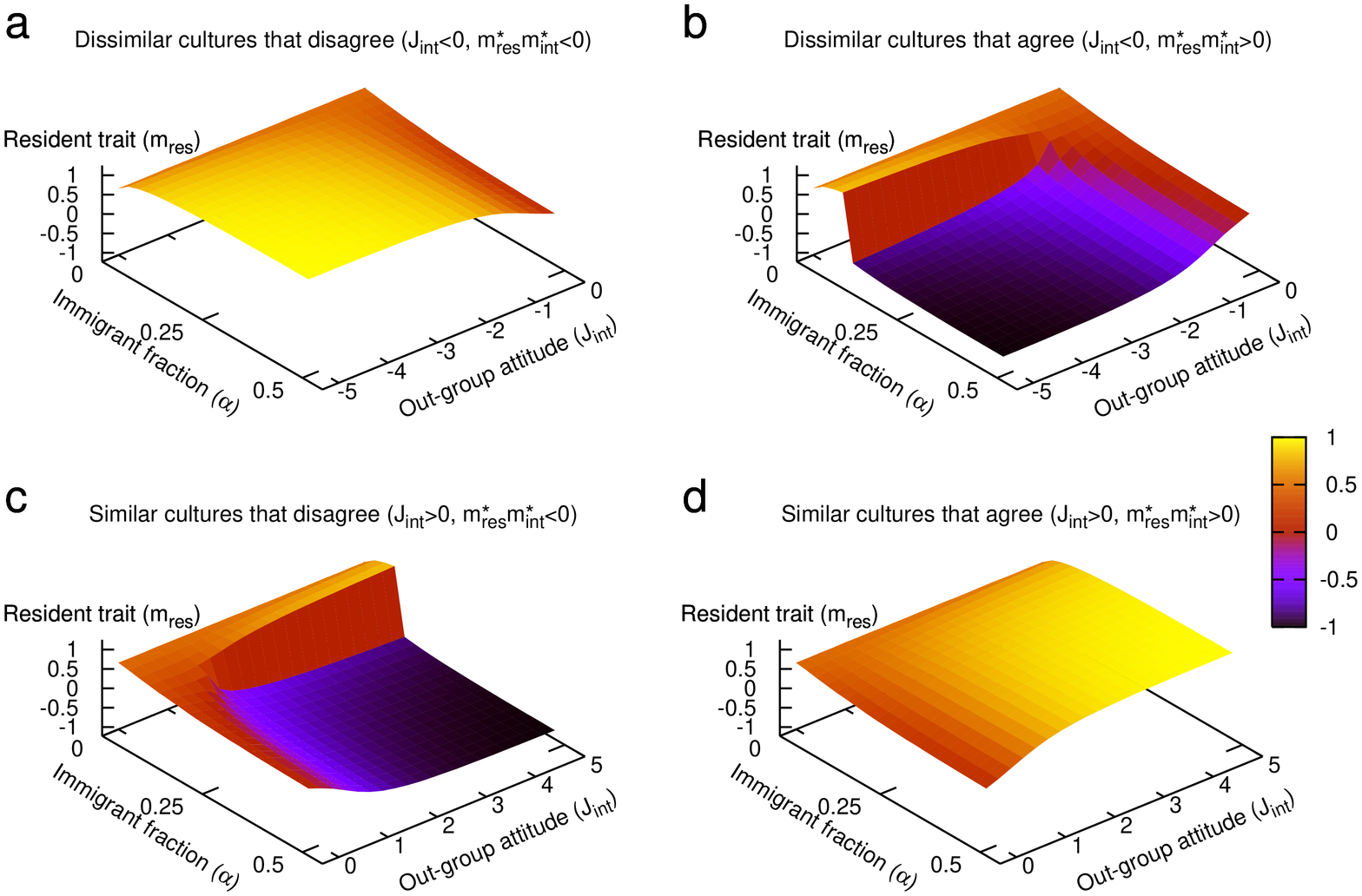}
  \caption{Influence of immigration on resident culture. Each panel
    portraits the mean resident trait after culture contact, \mres,
    as a function of the fraction of immigrants, $\alpha$, and the
    strength and sign of outgroup attitude, \Jint. Dramatic shifts in
    resident trait occur only when $\Jint\mres^*\mimm^*<0$
    (\textbf{b}, \textbf{d}) but not when $\Jint\mres^*\mimm^*$
    (\textbf{a}, \textbf{c}), where $\mres^*$ and $\mimm^*$ are,
    respectively, residents' and immigrants' mean traits before
    cultural interaction.  Parameter values: $\Jres=1$, $\Jimm=0.7$,
    $\mres^*=0.7$, $\mimm^*=-0.5$ (\textbf{a}, \textbf{c}) or
    $\mimm^*=0.5$ (\textbf{b}, \textbf{d}).}
  \label{fig:results}
\end{figure}
Numerical analysis of these equations reveals two main patterns of
behavior depending on the sign of the product $\mres^*\mimm^*\Jint$
(\fref{fig:results}).  When $\mres^*\mimm^*\Jint>0$ a small fraction
of immigrants causes only small changes in residents' trait, as
intuition would suggest. This includes two distinct cases: either the
two groups agreed before the interaction ($\mres^*\mimm^*>0$) and have
similar culture ($\Jint>0$, \fref{fig:results}a) or they disagree and
have dissimilar culture ($\mres^*\mimm^*<0$ and $\Jint<0$,
\fref{fig:results}d). The second pattern of results occurs when
$\mres^*\mimm^*\Jint<0$, in which case there exists a critical
fraction of immigrants, $\alpha_c$, above which residents suddenly
change to a nearly opposite mean trait value. This can happen either
when the two groups agree and have dissimilar culture
($\mres^*\mimm^*>0$ and $\Jint<0$, \fref{fig:results}b) or when the
groups disagree and have similar culture ($\mres^*\mimm^*<0$ and
$\Jint>0$, \fref{fig:results}c). This dramatic phenomenon only exists
when attitudes toward the outgroup (either positive,
\fref{fig:results}b, or negative, \fref{fig:results}c) are strong
enough. The shift can thus be inhibited by decreasing the magnitude of
\Jint. According to our assumptions, this amounts to making the groups
less similar when they are similar and less different when they are
different.

\begin{figure}[tp]
  \centering
  \includegraphics[angle=-90,width=.7\textwidth]{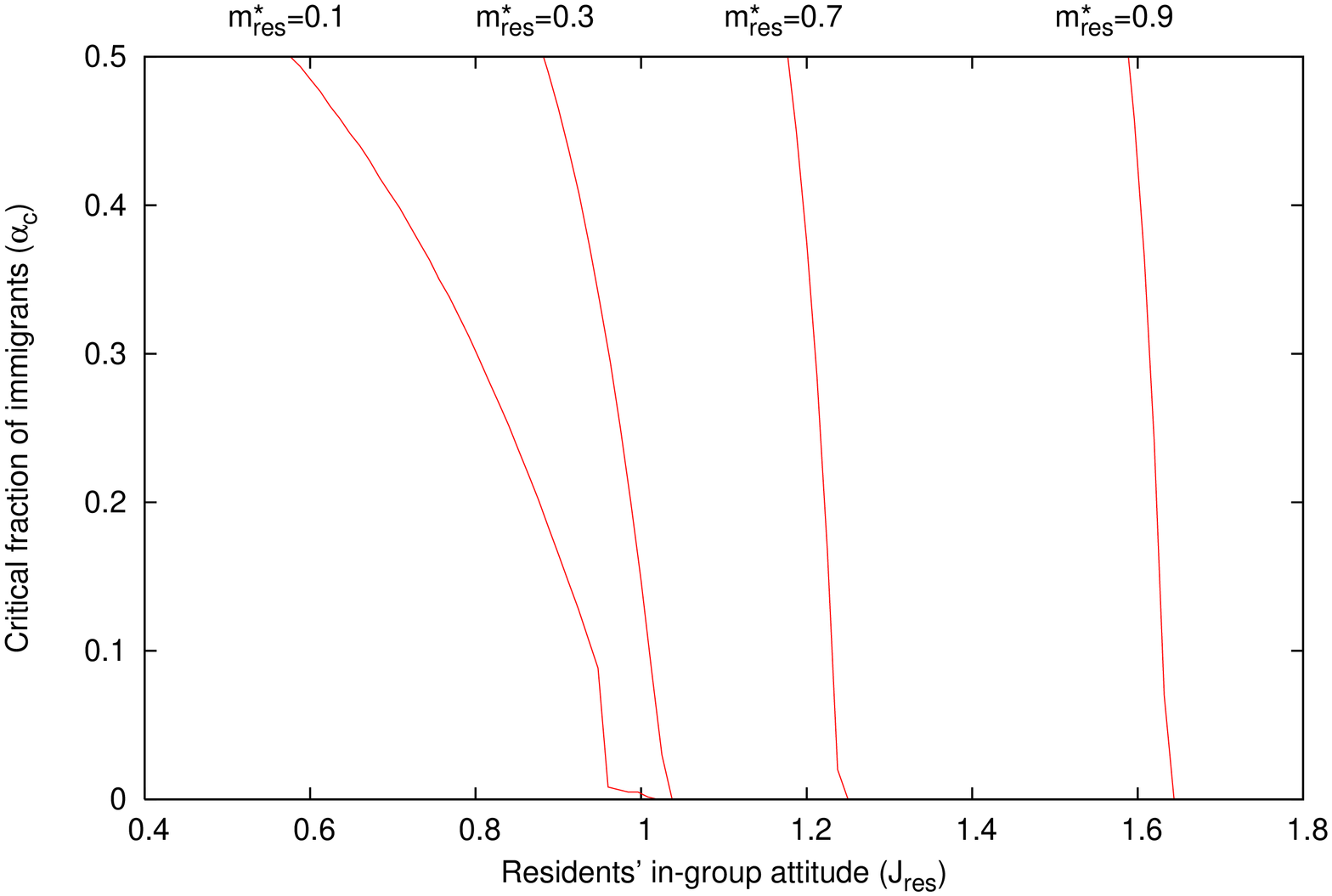}
  \caption{Ingroup attitudes and shifts in cultural traits. Each line
    shows the critical fraction of immigrants, $\alpha_c$, above which
    a sudden shift in residents' trait is observed, as a function of
    the strength of residents' ingroup attitude, $\Jres$. The curves
    are identical for the cases in \fref{fig:results}b and c, i.e.,
    for positive or negative $\Jint$.  For each value of $\mres^*$
    (different lines), lowering \Jres\ sharply increases the fraction
    of immigrants that can be sustained before residents significantly
    change trait.}
  \label{fig:alpha}
\end{figure}

\begin{figure}[tp]
  \begin{center}
    \begin{tabular}{@{}m{.495\textwidth}@{}m{.495\textwidth}@{}}
      \includegraphics[angle=-90,width=.45\textwidth]{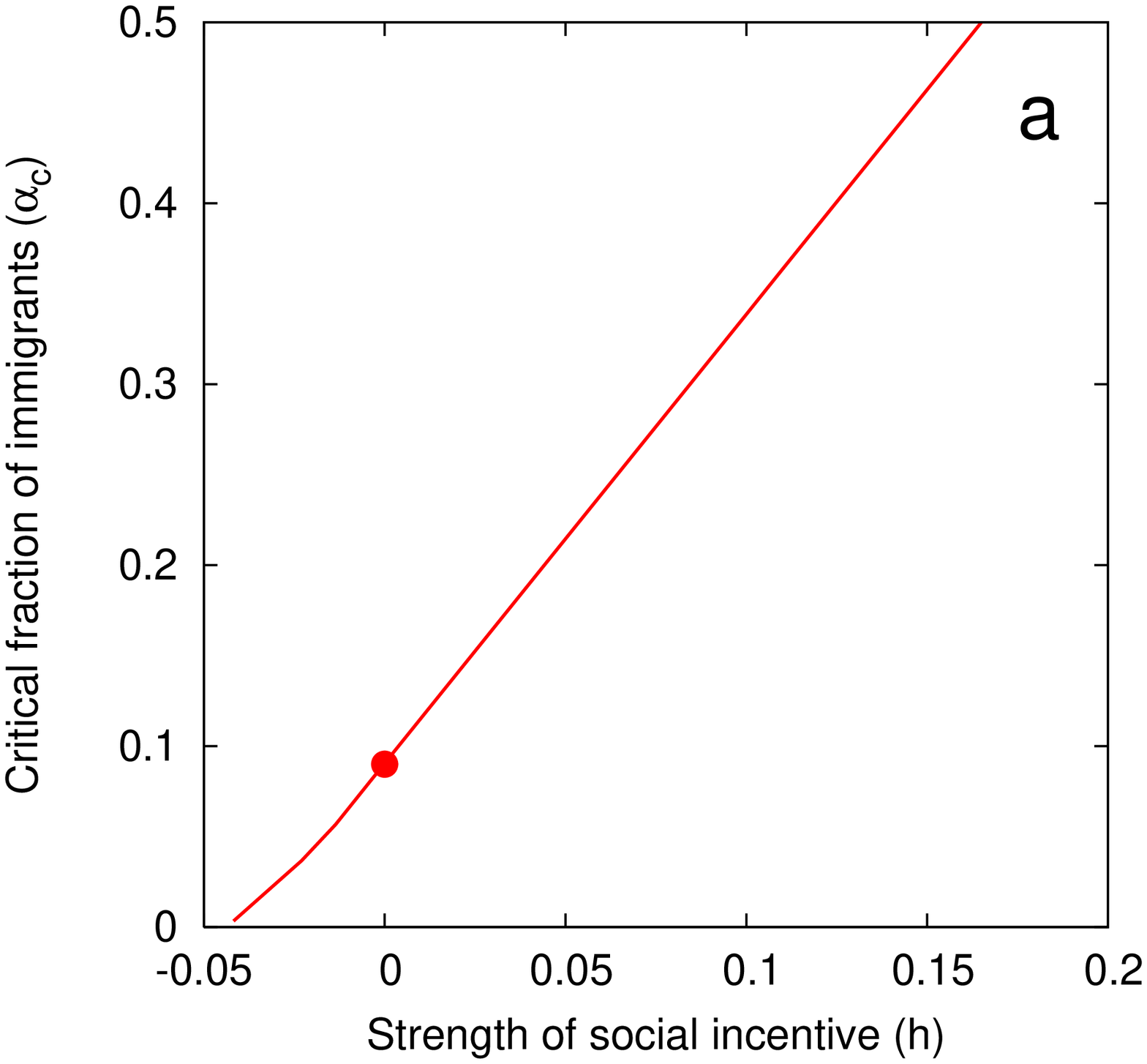} &    
      \includegraphics[angle=-90,width=.45\textwidth]{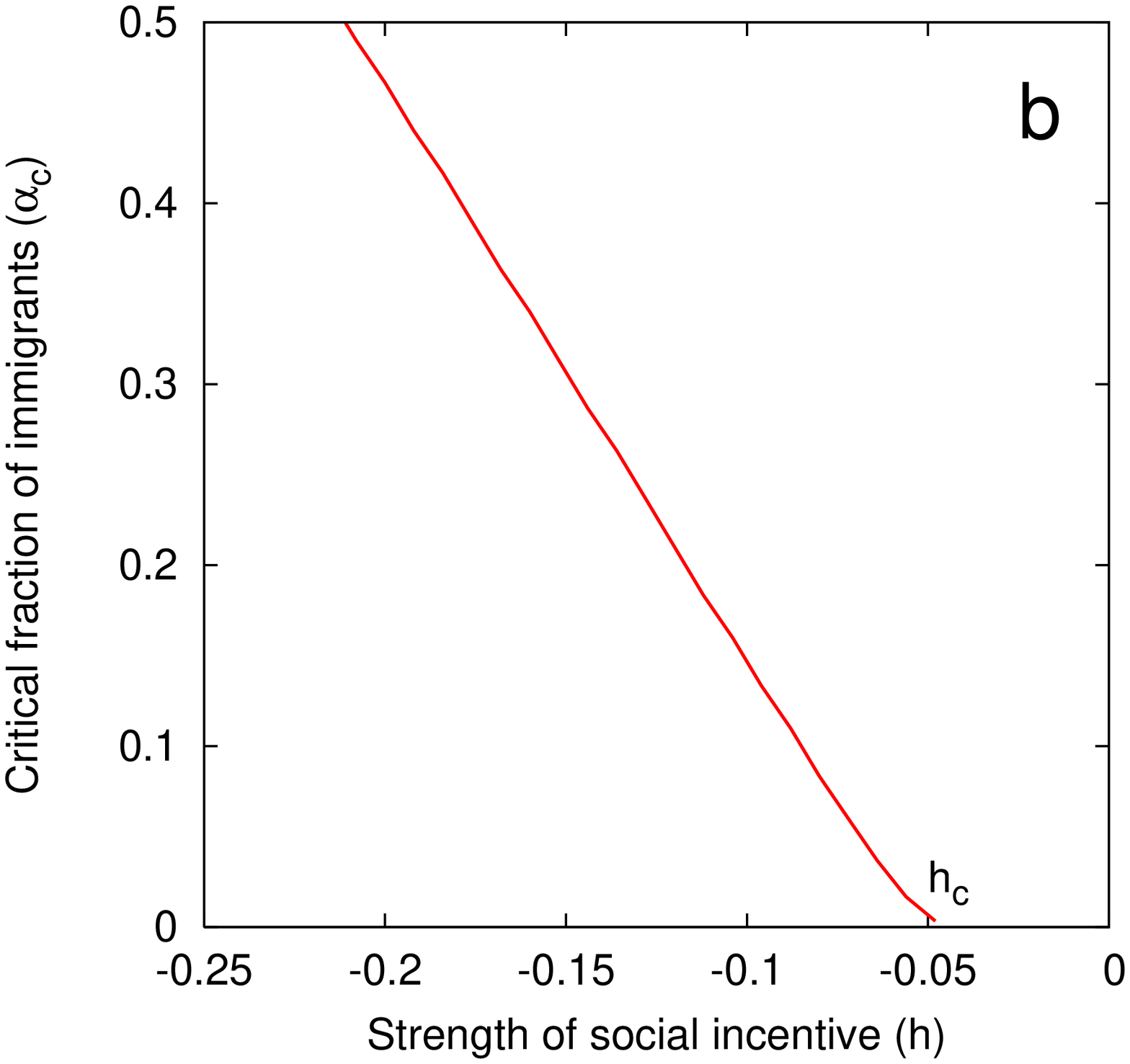}    
    \end{tabular}
  \end{center}
  \caption{Social incentives and shifts in cultural traits. Both panels
    plot the critical fraction of immigrants, $\alpha_c$, above which
    a sudden shift in residents' trait is observed, as a function of
    the strength of a social incentive capable of affecting individual
    traits, modeled as an external force, $h$ (equation \ref{eq:Hh}).
    \textbf{a}: a social incentive is added to the situation in
    \fref{fig:results}b ($h=0$, red dot in this figure).  If $h$
    favors (opposes) the residents' original trait value, the critical
    fraction of immigrants is raised (lowered). \textbf{b}: a social
    incentive is added to the situation in \fref{fig:results}d, in
    which no dramatic change is predicted.  When $h$ is decreased from
    0 toward negative values, residents' trait changes very slightly
    (not shown), until a critical value $h_c$ is reached when a sudden
    shift in residents' trait is predicted with a very small fraction
    of immigrants.  Parameter values as in \fref{fig:results}.}
  \label{fig:hext}
\end{figure}

There are other ways in which dramatic shifts in residents' trait may
be prevented.  It is possible, for instance, to reduce the strength of
residents' ingroup attitude, $\Jres$ (\fref{fig:alpha}).  According to
the similarity-attraction hypothesis, a higher value of \Jres\ 
corresponds to higher ingroup similarity. Hence our model suggests
that a more culturally homogeneous groups has a greater risk of
undergoing a dramatic transition when confronted with an immigrant
culture. That is, encouraging cultural diversity among residents may
make their culture more robust. One may also try to influence
individuals directly, introducing, e.g., social incentives that
encourage a given trait (modeled by $h$ terms as in equation
\ref{eq:Hh}).  Consider a situation in which residents are predicted
to change trait when the fraction of immigrants passes a critical
value, $\alpha_c$ (\fref{fig:results}b).  In such a case $\alpha_c$
can be increased subjecting individuals to an incentive $h$ that
favors the residents' original trait (\fref{fig:hext}a). An incentive
in the opposite direction, on the other hand, decreases $\alpha_c$,
suggesting that the impact of social policies can be dramatic. Indeed,
we show in \fref{fig:hext} that an incentive $h$ can trigger dramatic
changes even when these are impossible with $h=0$, as for instance in
\fref{fig:results}d.

\section{Discussion}

Clearly, our model is only an approximation to the complexities of
culture contact. Yet it exhibits a rich set of behaviors that, we
hope, may help to understand this complex phenomenon and promote the
development of more refined models. Our approach can be developed by
considering more individual traits (so that changes in culture as a
whole can be assessed), more realistic patterns of interactions among
individuals (social networks), individual variability in ingroup and
outgroup attitudes, and more complex rules of individual interaction.
Our basic hypothesis (similar individuals imitate each other more
strongly than dissimilar ones), however, is well rooted in social
psychology \citep{byrne71,byrn97}, including studies of intergroup
behavior \citep{grant93,mich02}, and we expect it to remain an
important feature of future models.

\section*{Acknowledgments}

We thank Francesco Guerra, Giannino Melotti and Magnus Enquist for
discussion. Research supported by the CULTAPTATION project of the
European Commission (FP6-2004-NEST-PATH-043434).

\bibliographystyle{plainnat} 
\bibliography{database,New}

\appendix

\section*{Appendix}

Statistical mechanics assigns to each system configuration $\s$ a
probability that is inversely related to the value of $H(s)$ through
an exponential function \citep{thompson79}
\begin{equation}
  \label{eq:Pr}
  \Pr(\s) = \frac{e^{-H(\s)}}{\sum_\s e^{-H(\s)}}
\end{equation}
where the denominator is a normalization factor ensuring
$\sum_s\Pr(\s)=1$. \Fref{eq:Pr} can be proved in a number of important
cases \citep{thompson79}, but is also used heuristically in more
general settings \citep{meza87}. In particular, the exponential is the
only function compatible with basic assumptions relevant for social as
well as physical systems, such that the probability of events relative
to two independent subsystems must multiply while other quantities
(e.g., entropy, see below) must add \citep{fermi36}.
 
The probability distribution (\ref{eq:Pr}) is used to calculate the
so-called \textit{free energy}, from which one can derive system-level
quantities such as the mean trait values \mres\ and \mimm. The free
energy is defined as the difference between \textit{internal energy}
$U$, i.e. the average of $H$ with respect to (\ref{eq:Pr}), and
\textit{entropy} $S=-\sum_\s\Pr\s\ln\Pr\s$:
\begin{equation}
  \label{eq:F}
  F=U-S
\end{equation}
For the model in the main text one may show that \citep{contucci_etal07}
\begin{equation}
\begin{array}{lll}
  U &=&  - \ \frac{1}{2}\big( \Jres(1-\alpha)^2\mres^2 + \Jimm\alpha^2\mimm ^2  +2\Jint \alpha(1-\alpha) \mimm \mres\big) +
         \\[5pt]  
    & &-  \  (1-\alpha) \hres^* \mres - \alpha \himm^* \mimm . 
   \end{array}
 \end{equation}
and
\begin{eqnarray}
S &=& (1-\alpha)\left( - \frac{1+\mres}{2}\ln \left(\frac{1+ \mres}{2}\right) - \frac{1-\mres}{2}\ln \left(\frac{1- 
\mres}{2}\right) \right) +
           \nonumber\\
  & &   \alpha \left(- \frac{1+\mimm}{2}\ln \left(\frac{1+ \mimm}{2}\right)  - \frac{1-\mimm}{2}\ln 
                \left(\frac{1- \mimm}{2}\right)\right).
        %       \nonumber
\end{eqnarray}
The values of \mres\ and \mimm\ are obtained from these expressions by
minimizing (\ref{eq:F}), which yields \fref{eq:m1m2}.

\end{document}